\RequirePackage{fix-cm}
\documentclass[smallextended]{svjour3}       
\smartqed  
\usepackage{graphicx}
 \usepackage{placeins}

\usepackage[
	n,
	advantage,
	operators,
	sets,
	adversary,
	landau,
	probability,
	notions,
	logic,
	ff,
	mm,
	primitives,
	events,
	complexity,
	asymptotics,
	keys
	]{cryptocode}

\usepackage{tikzfig}

\usepackage{stmaryrd}

\usepackage{slashbox}

\usepackage{tkz-graph}
\GraphInit[vstyle = Shade]
\tikzset{
  LabelStyle/.style = { rectangle, rounded corners, draw,
                        minimum width = 2em, fill = yellow!50,
                        text = red, font = \bfseries },
  VertexStyle/.append style = { inner sep=5pt,
                                font = \Large\bfseries},
  EdgeStyle/.append style = {->, bend left} }

\usepackage{lineno,hyperref}
\usepackage{algorithm}
\usepackage[noend]{algorithmic}
\usepackage{amssymb}
\usepackage{graphicx}
\usepackage{amsmath}
\usepackage{algorithm}
\usepackage[noend]{algorithmic}
\usepackage{url}
\usepackage{caption}
\usepackage{multirow}
\usepackage{threeparttable}
\usepackage{color}
\usepackage{lineno}

\usepackage{tkz-graph}

\tikzset{
  LabelStyle/.style = { rectangle, rounded corners, draw,
                        minimum width = 2em, fill = yellow!50,
                        text = red, font = \bfseries },
  VertexStyle/.append style = { inner sep=5pt,
                                font = \Large\bfseries},
  EdgeStyle/.append style = {->, bend left} }
  
  \usetikzlibrary{shapes.geometric}
\usetikzlibrary{positioning}

\begin{document}

\title{Quantum-enhanced Logic-based Blockchain \textrm{I}: Quantum Honest-success Byzantine Agreement and Qulogicoin 
}


\author{Xin Sun        \and
        Quanlong Wang
        \and
        Piotr Kulicki
        \and
        Xishun Zhao
}


\institute{Xin Sun \at
Institute of Logic and Cognition, Sun Yat-sen University, Guangzhou, China\\
Department of foundation of computer science, \\
The John Paul II Catholic University of Lublin, Lublin, Poland\\  
              \email{xin.sun.logic@gmail.com}           
           \and
           Quanlong Wang \at
            Department of computer science, \\ University of Oxford, Oxford, UK\\
           \email{harny3875@gmail.com}           
            \and
           Piotr Kulicki \at
           Department of foundation of computer science,  \\
           The John Paul II Catholic University of Lublin, Lublin, Poland\\
          \email{kulicki@kul.pl}   
          \and         
          Xishun Zhao \at
Institute of Logic and Cognition, Sun Yat-sen University, Guangzhou, China\\
              \email{hsszxs@mail.sysu.edu.cn}                  
}

\date{Received: date / Accepted: date}

\maketitle

\begin{abstract}

We proposed a framework of quantum-enhanced logic-based blockchain, which improves the efficiency and power of quantum-secured blockchain. The efficiency is improved by using a new quantum honest-success Byzantine agreement protocol to replace the classical Byzantine agreement protocol, while the power is improved by incorporating quantum protection and quantum certificate into the syntax of transactions. Our quantum-secured logic-based blockchain can already be implemented by the  current technology. The cryptocurrency created and transferred in our blockchain is called qulogicoin. Incorporating quantum protection and quantum certificates into blockchain makes it possible to use blockchain to overcome the limitations of some quantum cryptographic protocols. As an illustration, we show that a significant shortcoming of cheat-sensitive quantum bit commitment protocols can be overcome with the help of our blockchain and qulogicoin.

\end{abstract}

\section{Introduction}

A blockchain is a distributed ledger which enables achieving consensus in a large decentralized network of agents who do not trust each other. It is a ledger in the sense that data stored on a blockchain is transactions like ``Alice sends 1 bitcoin to Bob''.  It is distributed in the sense that every miners (the agents who are in charge of updating the ledger) has a same copy of the database. One of the most prominent application of blockchains is cryptocurrencies, such as Bitcoin \cite{Nakamoto08}. Another important application of blockchains is the implementation of self-executable smart contracts  \cite{Szabo97}. 

This is the first paper of our project on quantum-enhanced logic-based blockchain (QLB). The ultimate goal of our project is to design a framework of blockchain which has the following advantages over the existing classical and quantum blockchains:

\begin{enumerate}

\item More efficient.

\item More powerful.

\item More Secure.

\item Cheaper.

\item Smarter.

\item Easier-to-regulate.

\end{enumerate}
\noindent
In this paper we will use quantum technology to achieve efficiency, powerfulness, security  and cheapness. In the future, we will use techniques from logic to achieve smartness and regulatability.

Our starting point in this paper is the quantum-secured blockchain (QB) developed by Kiktenko \textit{et al.} \cite{Kiktenko17}.
Due to the application of quantum technologies, QB is more secure than classical blockchain in the sense that QB is immune from attacks of quantum computers, while classical blockchain is not. QB is probably also more efficient and (will be) cheaper than classical blockchain due to the omission of the costly and time-consuming proof-of-work. The security and cheapness of our QLB is inherited from QB. To achieve higher effiency, we will develop a new quantum Byzantine agreement (QBA) protocol to replace the classical Byzantine agreement protocol in QB. To make our blockchain more powerful, we will embed quantum protection and quantum certificate into the syntax of transactions in QLB. Those generalized transactions endow more power to QLB than QB and classical blockchain in the sense that QLB is able to handle contracts which involve tests of some quantum properties. As an illustration, we will show that a significant shortcoming of cheat-sensitive quantum bit commitment protocols can be overcome with the help of QLB, while classical blockchain and QB is unhelpful in this case.

The main contribution of this paper is the following:

\begin{enumerate}

\item We introduce QLB and demonstrate that it is more efficient and powerful than QB. Just like QB, QLB is also implementable by the current technology.

\item Based on QLB, a new cryptocurrency called qulogicoin is introduced.

\item Our QBA protocol is simpler and easier to be implemented than most QBA protocols in the literature. 

\item We discover a significant problem of quantum bit commitment and solve it. This problem is overlooked for many years.
\end{enumerate}

The structure of this paper is as follows. We present an overview of QLB in Section \ref{Overview of quantum-enhanced logic-based blockchain}. Then in Section \ref{A simplified quantum Byzantine agreement} we introduce a new QBA protocol to improve the efficiency of QLB. In Section \ref{Quantum logic and protected transaction} we demonstrate the power of QLB by solving a problem in quantum bit commitment. We draw conclusions in Section \ref{Conclusion and future work}.

\section{Overview of quantum-enhanced logic-based blockchain}\label{Overview of quantum-enhanced logic-based blockchain}

The structure of QLB is similar to the structure of QB \cite{Kiktenko17}.
We assume each pair of nodes (agents), of which at least a half of them are honest, is connected by an authenticated quantum channel and a not necessary authenticated classical channel.  
Each pair of agents can establish a sequence of secret keys by using quantum key distribution \cite{Bennett84}. Those keys will later be used for message authentication.

New transactions are created by those nodes who wish to transfer their cryptocurrency to another node. Each new transaction must contain the information about the Hash value, the receiver and the previous transaction from which the cryptocurrency is redeemed. Formally, a plain transaction $T_x$ saying that ``$i$ sends the qulogicoins, which $i$ has received from another transaction $T_y$, to $j$'' is of the form 
\begin{center}
$T_x=(x,  y, j)$.
\end{center} 
\noindent
Here $x$ is the Hash value of this transaction. A Hash function is a one-way function which maps an arbitrary length string to a fixed-length string. Just like in QB \cite{Kiktenko17}, we use Toeplitz hashing \cite{Krawczyk94,Krawczyk95}, of which the core component Toeplitz matrix is generated by the secret keys distributed via the quantum channel previously. Formally, $x= T_S (  y, j ) \oplus r$, where $S$ and $r$ are secret keys and $T_S$ is the Toeplitz matrix generated by $S$. $T_S()$ maps a string to another string of which the length is the same as the length of $r$. $\oplus$ is the exclusive-or operator.

A protected transaction $T_x$ saying that  `` $i$ sends the qulogicoins, which $i$ has received from another transaction $T_y$, to $j$; The qulogicoins  transferred in this transaction can only be used when both $\alpha$ and $ \phi$ is true'' is of the form 
\begin{center}
$T_x =(x, y, j;\alpha , \phi )$.
\end{center} 
Here $\alpha$ is a boolean function about the classical certificate and $\phi$ is a boolean function about the quantum certificate to be introduced soon.\footnote{In the future, we will use logical formulas to express $\alpha$ and $\phi$.} We let $x= T_S (  y, j;\alpha, \phi ) \oplus r$.

A more general form of a transaction is an extension of a protected transaction with some classical and quantum certificates. 

\begin{center}
$T_x =(x, y, j;\alpha , \phi ; \beta , \psi )$.
\end{center} 
\noindent 
Here $\beta$ is some classical data and $\psi$ is some quantum data, which are considered as certificates. We let $x= T_S (  y, j;\alpha, \phi; \beta ) \oplus r$. The functionality of certificates is to satisfy the protection condition of the the transaction $T_y$. An even more general form which involves more than 1 redeemed transaction $y_1,\ldots ,y_n$ can be defined as

\begin{center}
$T_x =(x, y_1, \ldots ,y_n, j;\alpha , \phi ; \beta_1 , \psi_1 ,\ldots ,\beta_n , \psi_n  )$.
\end{center} 

\noindent where $\beta_i , \psi_i$ are the classical and quantum certificates for $T_{y_i}$ and $x= T_S (  y_1,\ldots , $ $y_n, j;\alpha, \phi; \beta_1,\ldots \beta_n ) \oplus r$. 


A general transaction, except its quantum certificate, is then sent via classical channels to all miners, while the quantum certificate is sent via quantum channels. Each miner checks the consistency of the new transaction with respect to their local copy of the ledger 
and forms an opinion regarding the transaction's admissibility. Here consistency checking for $T_x =(x, y_1, \ldots ,y_n, j;\alpha , \phi ; \beta_1 , \psi_1 ,\ldots ,\beta_n , \psi_n  )$ means to check the following:

\begin{enumerate}
\item Message authentication: check if  $x= T_S (  y_1,\ldots , $ $y_n, j;\alpha, \phi; \beta_1,\ldots \beta_n ) \oplus r$, where $S$ and $r$ is taken from the secret keys shared between the miner and the sender.

\item check if the sender is the receiver of $T_{y_1} ,\ldots T_{y_n}$.

\item check if $T_{y_i}$ has been redeemed before this transaction, for all $i \in \{1,\ldots, n\}$.

\item check if $\beta_i$ satisfies $ \alpha_{y_i}$, where $\alpha_{y_i}$ is the classical protection of $T_{y_i}$.

\item check if $\psi_i$ satisfies $\phi_{y_i}$, where $\phi_{y_i}$ is the quantum protection of $T_{y_i}$.

\end{enumerate}

\noindent
Then all the miners apply the honest-success quantum Byzantine agreement protocol, which we will introduce in Section \ref{A simplified quantum Byzantine agreement}, to the new transaction, arriving at a consensus regarding the correct version of that transaction and whether the transaction is admissible. The double-spending events (a dishonest agent sending different versions of a particular transaction to different nodes of the network) is excluded in this stage. Finally, the transaction is added to the ledger of every node if at least a half of the miners agree that the transaction to be admissible. 

We will explain in the next section that some miners, as well as some special agents called list distributors, are rewarded in the procedure of achieving consensus. This is the only way to generate new cryptocurrency (qulogicoin) in our blockchain.

\section{Quantum honest-success Byzantine agreement}\label{A simplified quantum Byzantine agreement}

The Byzantine agreement protocol is the solution to the Byzantine generals problem  \cite{Pease80,Lamport82}:

\begin{center}

Three generals of the Byzantine army want to decide upon a common plan of action:
either to attack (0) or to retreat (1). They can only communicate in pairs by sending messages. However, one of the generals might be a traitor, trying to keep the loyal generals from agreeing on a plan. How to find a way
in which all loyal generals follow the same plan?

\end{center}


\begin{definition}[Byzantine agreement (BA) protocol \cite{Fitzi01}] A protocol among $n$ agents such that one distinct agent $S$ (the sender) holds an input value
$x_s \in D$ (for some finite domain $D$) and all other agents (the receivers) eventually decide on an output value in $D$ is said to achieve Byzantine agreement if the protocol guarantees that all honest agents decide on the same output value $y\in D$ and that $y=x_s$ whenever the sender is honest.

\end{definition}






In QB \cite{Kiktenko17}, the authors use classical Byzantine agreement protocol  \cite{Pease80} to update the distributed ledger. They noticed that a shortcoming of the classical Byzantine agreement protocol \cite{Pease80} is that it becomes exponentially data-intensive if a large number of cheating nodes
are present. Therefore further research on developing an efficient consensus protocol is required.

In the literature of quantum computing,  several Byzantine agreement protocols has been studied in the past decade \cite{Fitzi01,Iblisdir04,BenOr05,Gaertner08,Tavakoli15}.
In the setting of QLB, we need a Byzantine agreement protocol to solve the double-spending problem. It turns out that the following weak notion of Byzantine agreement is already sufficient for our purpose. 
\begin{definition}[honest-success Byzantine agreement protocol (HBA)] A protocol among $n$ agents such that one distinct agent $S$ (the sender) holds an input value
$x_s \in D$ (for some finite domain $D$) and all other agents (the receivers) eventually decide on an output value in $D$ is said to achieve honest-success Byzantine agreement if the protocol guarantees the following: 

\begin{enumerate}
\item If the sender is honest, then all honest agents decide on the same output value $y=x_s$.

\item If the sender is dishonest, then either  all honest agents abort the protocol, or all honest agents decide on the same output value $y\in D$.

\end{enumerate}

\end{definition}

We say that a HBA protocol is $p$-resilient, where $0<p<1$, if the protocol still works when less than a fraction of $p$ receivers are dishonest. The quantum honest-success Byzantine agreement (QHBA) protocol that we will present in this section is $\frac{m-2}{m}$--resilient, where $m$ is the number of receivers. Our protocol is much more efficient than classical BA protocol in the presence of a large number of cheating nodes. 

There are three phases of our QHBA  protocol. The aim of the first phase is to distribute a set of correlated lists among agents. In the second phase, some special correlated lists are generated based on the set of correlated lists. Then in the third phase, agents use the special correlated lists to achieve consensus.

\subsection{List distribution by quantum secure direct communication}

Unlike quantum key distribution, which only allows to distribute non-deterministic message, quantum secure direct communication (QSDC) \cite{Chamoli09,Naseri10,Naseri15} allows messages to be deterministically sent through the quantum channel. We use QSDC to distribute those correlated lists. Our QSDC protocol is based on a quantum version of Shamir's three-pass protocol \cite{Zhou18}.   A classical scenario of the three-pass protocol is the following \footnote{\url{https://en.wikipedia.org/wiki/Three-pass_protocol}}
\begin{itemize}
\item Alice's friend Bob lives in a repressive country where the police spy on everything and open all the mails.
\item Alice needs to send a valuable object to Bob.
\item Alice has a strongbox with a hasp big enough for several
locks, but no lock to which Bob also has a key.

\end{itemize}
\noindent
How can Alice get the item to Bob securely? Alice and Bob might take the following three pass protocol:

\begin{enumerate}

\item Put the item into the box, attach Alice's lock to the hasp, and mail the box to Bob.

\item Bob adds his own lock and mails the box back to Alice.

\item Alice removes her lock and mails the box back to Bob. Bob now removes his lock and opens the box.
\end{enumerate}

In a previous paper \cite{Zhou18}, we have introduced a quantum realization of the three-pass protocol for key distribution. It turns out that this protocol can be straightforwardly used for secure direct communication. Now we recap the quantum three-pass protocol in \cite{Zhou18}.

We use qubit $|0\rangle $ and $|1\rangle $ to encode 0 and 1 respectively. Our key space for encryption and decryption contains 4 X-gates $\{X(0),X(\frac{\pi}{2}),X(\pi), X(\frac{3\pi}{2})\}$, where $X(m)= |+ \rangle \langle +| +e^{mi}  |-\rangle \langle -| $. The encryption of a qubit $|i\rangle$  with key $k$ is defined as $Enc_{k}(i) = k |i\rangle $, and the decryption of a qubit $|i\rangle$ is $Dec_{k}(i) = k |i\rangle $, where $k \in \{X(0),X(\frac{\pi}{2}),X(\pi), X(\frac{3\pi}{2})\}$. We let $(X(m),\overline{X(m)}=X(2\pi -m))$ be a pair of encryption/decryption keys.

\begin{figure}
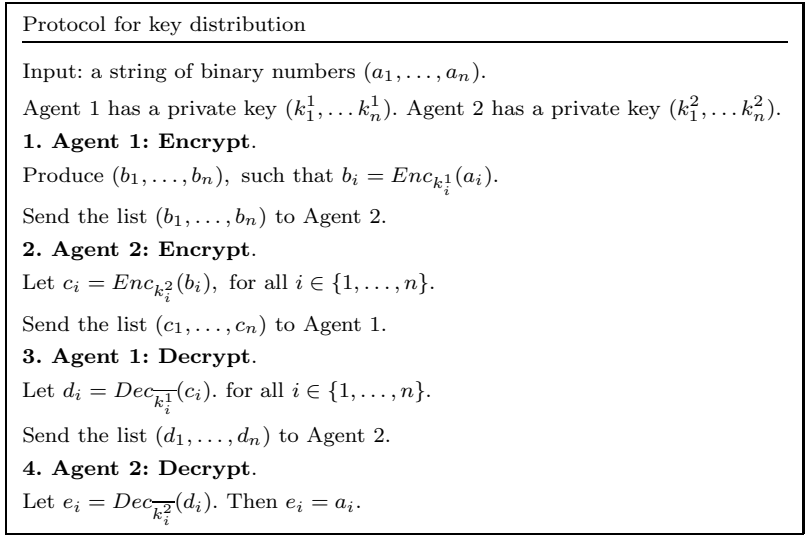


\begin{center}
\fbox{%
\procedure{Protocol for key distribution}{%
\text{Input: a string of binary numbers }
(a_1, \ldots, a_n ). \\
\text{Agent 1 has a private key } (k^1_1,\ldots k^1_n). \text{ Agent 2 has a private key } (k^2_1,\ldots k^2_n).\\
         \textbf{1. Agent 1: Encrypt}.  \\
\text{Produce }(b_1, \ldots, b_n ),
\text{ such that  } b_i = Enc_{k^1_i}(a_i).\\
\text{Send the list }(b_1, \ldots, b_n ) \text{ to Agent } 2.\\
           \textbf{2. Agent 2: Encrypt}. \\
\text{Let }c_i = Enc_{k^2_i} (b_i ), \text{ for all } i\in \{1,\ldots,n\}.\\
\text{Send the list }(c_1, \ldots, c_n )  \text{ to Agent } 1.\\
         \textbf{3. Agent 1: Decrypt}. \\
\text{Let }d_i = Dec_{\overline{k^1_i}} (c_i ).  \text{ for all }  i \in \{1, \ldots, n\}.\\ \text{Send the list }(d_1,\ldots, d_n ) \text{ to Agent } 2.\\
      \textbf{4. Agent 2:  Decrypt}. \\
\text{Let }e_i = Dec_{\overline{k^2_i}} (d_i ).  \text{ Then }e_i = a_i.
  }
}
\end{center}

\caption{A quantum three-pass protocol for secure direct communication}
		\label{A protocol for key distribution}
	
\end{figure}

Figure \ref{A protocol for key distribution} is our quantum three-pass protocol for a sender (agent 1) to send a sequence of bits  to a receiver (agent 2). At the beginning of the protocol, agent 1
encrypts the bit string element-wise and sends the resulting string to agent 2. Then agent 2  encrypts the ciphertext and sends the result back to agent 1. Agent 1 then decrypts the string and sends it to agent 2. Now agent 2 decrypts the string and gets the key.

Now we use our quantum three-pass protocol to distribute correlated lists. Let $\{P_1, \ldots , P_n, P_{n+1}, \ldots , P_{n+d}\}$ be the set of agents. We let $P_1$ be the sender of the QHBA protocol, $P_2, \ldots , P_{n}$  be receivers and $P_{n+1}, \ldots , P_{n+d}$ be list distributors who are in charge of distributing lists of correlated numbers. For every agent $P_i \in \{P_{n+1}, \ldots , P_{n+d}\}$, the task of $P_i$ is to use the quantum three-pass protocol to send a list of numbers $L^i_k$ to agent $P_k \in \{P_1, \ldots , P_{n}\}$ such that the following is satisfied:
\begin{enumerate}

\item For all $k\in \{1,\ldots,n\}$, $|L^i_k | =m$ for some integer $m$ which is a multiple of 6.
\item $L^i_1 \in \{0,1,2\}^m$. $\frac{m}{3}$ numbers on $L^i_1$ are $0$.  $\frac{m}{3}$ numbers on $L^i_1$ are $1$.  $\frac{m}{3}$ numbers on $L^i_1$ are $2$.
\item For all $k\in \{2,\ldots,n\}$, $L^i_k \in \{0,1\}^m$.
\item For all $j\in \{1,\ldots,m \}$, if $L^i_1[j]=0$, then $L^i_2[j]=\ldots= L^i_{n}[j] =0$.
\item For all $j\in \{1,\ldots,m \}$, if $L^i_1[j]=1$, then $L^i_2[j]=\ldots= L^i_{n}[j] =1$.

\item For all $j\in \{1,\ldots,m \}$, if $L^i_1[j]=2$, then for all $k\in \{2, \ldots ,n \}$ the probability of $L^i_k[j]=0$ and $L^i_k[j]=1$ are equal.

\end{enumerate}

After those lists are selected\footnote{Some additional decoy qubits may be sent and revealed to detect eavesdroppers.}, $P_2,\ldots P_n$ are entitled to communicate with  $P_1$ to check whether those lists satisfy the above specification. If for some parameter $\theta \in [0,\frac{1}{2}]$, more than $\theta n$ agents report that the lists distributed by $P_i$ do not satisfy the specification. The $P_i$ is classified as a corrupted/dishonest distributor. To stimulate $P_i$ to play honestly, honest distributors will be rewarded by some qulogicoins while those corrupted distributors will not.


\subsection{List formation by sequential composition}

In this phase, $\{P_1, \ldots, P_n\}$ use a simply sequential composition procedure to form a unique list to be used in the next phase.

Assume there are $h$ list distributors who are classified to be honest. Without loss of generality, the lists distributed by those distributors $P_{n+1}, \ldots, P_{n+h}$
can be represented by $\mathfrak{L}^{n+1} =( L^{n+1}_1 ,\ldots ,L^{n+1}_n), \ldots, \mathfrak{L}^{n+h} =( L^{n+h}_1 ,\ldots ,L^{n+h}_n)$. The aim of this phase is to form a new sequence of lists $\mathfrak{L} =( L_1 ,\ldots ,L_n)$, which is to be used in the next phase. We construct $\mathfrak{L}$ by the sequential composition of $\mathfrak{L}^{n+1},\ldots ,\mathfrak{L}^{n+h}$. That is, we let $L_1= L_{1}^{n+1}\ldots L_{1}^{n+h}, \ldots ,L_n= L_{n}^{n+1}\ldots L_{n}^{n+h} $.

On average, every honest distributor contributes $\frac{1}{h}$ to the final lists $\mathfrak{L}$. This property is crucial to counteract the attack of some adversary who tries to bribe the distributor. We will briefly discuss this attack in Section \ref{Some potential attacks}. This mechanism together with the mechanism of reward will encourage more distributors to behave honestly. On the other hand, the more honest distributors there are, the more reliable is our blockchain.

\subsection{Achieving consensus}

After the correlated lists $\mathfrak{L} =( L_1 ,\ldots ,L_n)$ are established,
the agents $P_1,\ldots , P_n$, of which we assume at least a half of them are honest, run the following steps to achieve consensus. 

\begin{enumerate}
\item $P_1$ sends a binary number $b_{1,k}$ to all $P_k$, $k\in \{2,\ldots,n\}$. Together with $b_{1,k}$, $P_1$ sends a list of numbers $ID_{1,k}$, which indicates all positions on $L_1$ where $b_{1,k}$ appears, to $P_k$. The length of $ID_{1,k}$ is required to be $\frac{m}{3}$, where $m$ is the length of $L_1$. If $P_1$ is honest, then he will send the same message to all agents, \textit{i.e.} $(b_{1,k},ID_{1,k} ) = (b_{1,j},ID_{1,j} )$ for all $j,k \in \{2,\ldots,n\}$. If $P_1$ is dishonest, then he will send different binary numbers and different lists of numbers to different agents, \textit{i.e.} $(b_{1,k},ID_{1,k} ) \neq (b_{1,j},ID_{1,j} )$ for some $j,k$.

An honest $P_1$ will also use $b_{1,k}$ as the final value it outputs, while a dishonest $P_1$ will use $b_{1,k}$ or $1- b_{1,k}$ randomly as its final output value.

\item Agent $P_k$ analyzes the obtained message $(b_{1,k}, ID_{1,k})$ with his own list $L_k$. If the analysis of $P_k$ shows that $(b_{1,k},ID_{1,k} ) $ is
consistent with $L_k$, then if $P_k$ is honest, he 
sends $(b_{1,k},ID_{1,k} ) $ to all other agents  $P_{j \neq 1}$. Here $(b_{1,k},ID_{1,k} ) $ is
consistent with $L_k$ means that for all index $x\in ID_{1,k}$, $L_k[x]=b_{1,k}$. However, if $(b_{k,j},ID_{k,j} ) $ is not consistent
with $L_k$, then $P_k$ immediately ascertains that $P_1$ is dishonest and
sends $\bot$ to other agents, meaning that ``I have received inconsistent message''. A dishonest $P_k$ sends $1- b_{1,k}$ with whatever indexes he chooses or simply $\bot$. The full information which $P_j$ receives from $P_k$ will be denoted by $(b_{k,j},ID_{k,j} ) $.


\item After all messages have been exchanged between $P_2,\ldots , 	P_n$, every honest agent $P_k$ considers the received data and acts according to the following criterion:

\begin{enumerate}

\item If there is a set of agents $H$ with $|H|\geq 2$ such that 
 
\begin{enumerate}
\item for all $j\in H$, $(b_{j,k}, ID_{j,k})$ is consistent with $L_k$, and 
\item for some $i,j \in H$, $b_{i,k} \neq b_{j,k}$, 

\end{enumerate} 
then $P_k$ sets his output value to be $\bot$.

 \item If there is a set of agents $H$ with $|H|\geq 2$ such that for all $j\in H$, $(b_{j,k}, ID_{j,k})$ is consistent with $L_k$ and all $b_{j,k} $ are the same, and for all  $i \not\in H$, $(b_{i,k}, ID_{i,k})$ is not consistent with $L_k$, then $H$ is the set of all honest agents and $P_k$ sets his output value $v_k = b_{j,k}$.

 \item If there is a set of agents $H$ with $|H|\geq 2$ such that for all $j\in H$, $(b_{j,k}, ID_{j,k})$ is consistent with $L_k$ and all $b_{j,k} $ are the same, and for all  $i \not\in H$, the message sent by $P_i$ is $\bot$, then $P_k$ set $v_k = b_{j,k}$.

\item In all other cases, $P_k$ sets his value to be $\bot$.

\end{enumerate}

\item We say that consesus is achieved if at least $\frac{n}{2}$ agents output the same bit value $v\in \{0,1\}$. In this case, those agents whose output is the same as $v$ are rewarded with  some qulogicoins.

\end{enumerate}

Now we briefly explain the rationale of these criterion. Suppose $P_j$ is a dishonest receiver, where $j\geq 2$. Now $P_j$ wants to send $(b_{j,k}, ID_{j,k})$ to $P_k$ such that  $(b_{j,k}, ID_{j,k})$ is consistent with $L_k$.
Note that on $L_j$, there are $\frac{m}{2}$ positions on which $b_{j,k}$ appears. But on $L_1$, there are only $\frac{m}{3}$ positions on which $b_{j,k}$ appears. Therefore, there are $\frac{m}{2}-\frac{m}{3}= \frac{m}{6}$ positions on which there are some discord. But $P_j$ has no knowledge about those discord. We say a position $x$ is a discord position iff $L_1[x]=2$. If $P_j$ selects a discord position $x$ and put it into $ID_{j,k}$, then with probability $\frac{1}{2}$ it will be that $L_k[x] \neq b_{j,k}$. Therefore, to ensure that $(b_{j,k}, ID_{j,k})$ is consistent with $L_k$, $P_j$ has to avoid all discord positions. The probability of avoiding all discord positions is $(\frac{2}{3})^{\frac{m}{3}}$, which is extremely small when $m$ is relatively large. Therefore, if it happens that  
$(b_{j,k}, ID_{j,k})$ is consistent with $L_k$, then $P_k$ can conclude that $P_j$ is honest. For the same reason $P_k$ can conclude that $P_i$ is honest when $(b_{i,k}, ID_{i,k})$ is consistent with $L_k$. Now, if in addition that 
$b_{i,k} \neq b_{j,k}$, $P_k$ can safely conclude that $P_1$ is dishonest. This is the rationale of criterion (a).

The rationale of criterion (b) is essentially the same as the rationale of criterion (a). In this case, $P_1 \in H$ and those agents who are not in $H$ are dishonest agents and they failed on cheating.

The rationale of criterion (c) is also essentially the same as the rationale of criterion (a). In this case some agents who are not in $H$ is also honest. They will change their output value from $\bot$ to $b_{j,k}$.

\subsection{Some potential attacks}\label{Some potential attacks}

Now we briefly discuss some potential attacks to our blockchain.

\subsubsection{Double-spending adversary}

Our blockchain is resistant to a double-spending adversary. A double spending adversary can simply be modeled by a dishonest sender $P_1$ who sends $0$ to some nodes and $1$ to some other nodes. This kind of attack will not work according to criterion (a).

\subsubsection{Tampering adversary}

Our blockchain is also resistant to a tampering adversary. We model a tampering adversary by an arbitrary set of dishonest receivers, of which the cardinality is less then $\frac{n}{2}$. A dishonest receiver $P_i$ maliciously sends $1- b_{1,i}$, together with some indexes, to all other agents. This kind of adversary will not be successful according to criterion (b) and (c), as long as there are still 2 honest receivers.

\subsubsection{Bribe the distributor}

  Every honest list distributor has some knowledge about the list $\mathfrak{L} =( L_1 ,\ldots ,L_n)$. Some adversary may bribe list distributors to obtain some valuable knowledge about $\mathfrak{L}$. But this attack is costly for the adversary, because he has to bribe many list distributors to obtain sufficient knowledge.

\section{Application in quantum bit commitment} \label{Quantum logic and protected transaction}

Now we apply our QLB to the design of quantum bit commitment protocols. Bit commitment, used in a wide range of cryptographic protocols (e.g. zero-knowledge proof, multiparty secure computation, and oblivious transfer), typically consists of two phases, namely: commitment and opening.
In the commitment phase, Alice the sender chooses a bit $a$ ($a = 0$ or $1$) which she wishes to commit to the receiver Bob. Then Alice presents Bob some evidence about the bit. The committed bit cannot be known by Bob prior to the opening phase. Later, in the opening phase, Alice announces some information for reconstructing $a$. Bob then reconstructs a bit $a'$ using Alice's evidence and announcement. A correct bit commitment protocol will ensure that $a' = a$. A bit commitment protocol is concealing if Bob cannot know the bit Alice committed before the opening phase and it is binding if Alice cannot change the bit she committed after the commitment phase. 

The first quantum bit commitment (QBC) protocol is proposed by Bennett and Brassard in 1984 \cite{Bennett84}. A QBC protocol is unconditionally secure if any cheating can be detected with a probability arbitrarily close to 1. Here, Alice's cheating means that Alice changes the committed bit after the commit phase, while Bob's cheating means that Bob learns the committed bit before the opening phase. A number of QBC protocols are designed to achieve unconditional security, such as those of \cite{Brassard90,Brassard93}. However, according to the Mayers-Lo-Chau (MLC) no-go theorem \cite{Mayers97,LoChau97}, unconditionally secure QBC can never be achieved in principle.

Although unconditional secure QBC is impossible, several QBC protocols satisfy some other notions of security, such as cheat-sensitive quantum bit commitment (CSQBC) protocols \cite{Hardy04,Buhrman08,Shimizu11,Li14,ZhouSun18} and relativistic QBC protocols \cite{Kent11,Adlam15}. In CSQBC protocols, the probability for detecting cheating is merely required to be non-zero. With this less stringent security requirement, many  QBC protocols which are not unconditional secure are regarded as secure in the cheat-sensitive sense. 

However, for any CSQBC to work, there has to be a mechanism of enforcing punishment when a cheating behavior is detected. Otherwise both Alice and Bob will always cheat, regardless of whether it will be detected. Therefore the protocol will either be aborted or end with someone cheats  successfully. So far this problem (enforcing punishment in CSQBC) is completely omitted in the literature. In this section, we solve this problem by applying QLB to enforce punishment.\footnote{While the Bitcoin blockchain has been applied to solve a related problem in classical bit commitment \cite{Andrychowicz16}, the solution seems to be inadequate for quantum bit commitment because there is no quantum data in the Bitcoin blockchain.}

\subsection{A  cheat-sensitive quantum bit commitment protocol}
For the sake of concreteness, we present a CSQBC protocol taken from a recent paper of ours \cite{SunWang18} and demonstrate how to enforce punishment for this protocol by using QLB. There are three phases of this protocol, namely: the preparation phase, the commitment phase and the opening phase. 

The preparation phase contains the following steps:

\begin{enumerate}

\item Bob generates a sequence of $n$ qubits such that
\begin{enumerate}
\item $n$ is a multiple of 4.

\item  $\frac{n}{4}$ qubits are $|0 \rangle$,  $\frac{n}{4}$ qubits are  $X(\frac{\pi}{2})|0 \rangle = |i \rangle$,  $\frac{n}{4}$ qubits are  $X|0 \rangle = |1 \rangle$, and  $\frac{n}{4}$ qubits are $X(\frac{3\pi}{2})|0 \rangle =| \overline{i} \rangle$.

\item for every $j\in \{1,\ldots , \frac{n}{2}\}$, the $2j-1$th qubit and the $2j$th qubit are from different ONBs. 

\end{enumerate}
Such a sequence is called a \textit{balanced-uniform} sequence.  Bob generates $m$ balanced-uniform sequences and sends them to Alice.

\item Alice chooses $m-1$ sequences and asks Bob to reveal, qubit by qubit, which state it was prepared. Then, Alice measures those qubits in the appropriate basis to verify whether Bob  has prepared those qubits in the required specification: the $\{ |0 \rangle, |1 \rangle \}$ basis for qubits $|0 \rangle$ and $|1 \rangle$ and the $\{ |i \rangle, |\overline{i} \rangle \}$ basis for qubits $|i \rangle$ and $|\overline{i} \rangle$. If Alice  detects that Bob  has prepared a sequence that is not balanced-uniform, then Alice has detected Bob's cheating.

\end{enumerate}

The commitment phase contains the following steps:

\begin{enumerate}

\item Alice commits 2 bits by applying quantum operations to the only balanced-uniform sequence left. We denote this sequence as $QS$. If Alice decides to commit $00/01/10/11$, then he/she applies $X(0)/X(\frac{\pi}{2})/X(\pi)/X(\frac{3\pi}{2})$ to all qubits in $QS$, respectively. Then, Alice generates a classical string $CS$ of length $\frac{n}{2}$. Alice applies the SWAP operator to $QS[2j-1]$ and $QS[2j]$ iff $CS[j]=1$. Alice sends $QS$ to Bob.

\item Bob measures each received qubit either in the $\{ |0 \rangle, |1 \rangle \}$ basis
or in the $\{ |i \rangle, |\overline{i} \rangle \}$ basis which is chosen uniformly at random.

\end{enumerate}

The opening phase contains the following steps:

\begin{enumerate}
\item Alice reveals $CS$.

\item Based on the information of $CS$, Bob is able to know the original state of each position in QS, because it is Bob who prepared QS and now he knows how it was swapped.
Now Bob determines whether it was measured in the correct basis for
each position in QS: for a position that was originally occupied by $|0\rangle $ or $|1\rangle $, the correct basis is the $\{|0\rangle, |1\rangle\}$ basis, for other qubits the correct basis is the other basis.
Now, Bob can reconstruct the bits committed by Alice as follows:

\begin{enumerate}
\item If Alice committed to $00$, then all the qubits measured in the correct
basis must yield a state which is the same as the original one. 
\item If Alice committed to $10$, then all the qubits measured in the correct
basis must yield a state which can be recovered to the original one by applying a $X(\pi)$ gate afterwrds. 

\item If Alice committed to $01$, then all the qubits measured in the incorrect
basis must yield a state which can be recovered to the original one by applying a $X(\frac{3\pi}{2})$ gate afterwrds. 


\item If Alice committed to $11$, then all the qubits measured in the incorrect
basis must yield  a state which can be recovered to the original one by applying a $X(\frac{\pi}{2})$ gate afterwrds. 


\end{enumerate}

All other cases are classified as Alice's cheating.

\end{enumerate}

\subsection{Enforce punishment by QLB}

To enforce punishment for Bob when he cheats, we require Bob to create the following protected transaction on QLB. ``Bob sends $n$ qulogicoins, which he has received from another transaction $T_y$, to Bob; The qulogicoins transferred in this transaction can only be used when the following is true:
 
 \begin{itemize}
\item   Alice discover that all sequence prepared by Bob are balanced-uniform. Or equivalently, Alice does not discover Bob's cheating in the preparation phase.''
 
 \end{itemize}

To enforce punishment for Alice when she cheats, we require Alice to create the following protected transaction on QLB. ``Alice sends $n$ qulogicoins, which she has received from another transaction $T_y$, to Alice; The qulogicoins transferred in this transaction can only be used when the following is true:
 
 \begin{itemize}
\item Bob does not discover Alice's cheating in the opening phase. Or equivalently, one of the following is true:

\begin{enumerate}
\item All the qubits measured in the correct
basis yield a state which is the same as the original one. 
\item All the qubits measured in the correct
basis yield a state which can be recovered to the original one by applying a $X(\pi)$ gate. 

\item All the qubits measured in the incorrect
basis yield a state which can be recovered to the original one by applying a $X(\frac{3\pi}{2})$ gate. 


\item All the qubits measured in the incorrect
basis yield a state which can be recovered to the original one by applying a $X(\frac{\pi}{2})$ gate.''\footnote{Although these quantum protections is expressed informally in this paper, they can be expressed concisely and precisely by some quantum logic, such as the the dynamic logic of quantum programs \cite{Baltag05,Baltag06,Baltag08,Baltag14,Bergfeld17}. We leave this as future work.} 


\end{enumerate}
 
 \end{itemize}

\section{Conclusion and future work}\label{Conclusion and future work}

In this paper we introduced quantum-enhanced logic-based blockchain to improve the efficiency and power of quantum-secured blockchain \cite{Kiktenko17}. The efficiency is improved by using a new quantum honest-success Byzantine agreement protocol to replace the classical Byzantine agreement protocol in quantum-secured blockchain, while the power is improved by incorporating quantum certificate and quantum protection into the syntax of transactions.
No multi-particle entanglement is used in our blockchain, which makes it easy to be implemented  with the current technology. In fact, all the quantum technology needed for our blockchain is already available in laboratories and even industry. Incorporating quantum certificate and quantum protection into blockchain makes it possible to use our blockchain to overcome a significant shortcoming of cheat-sensitive quantum bit commitment protocols. 

In the future, we will further extend QLB to make it smarter and easier-to-regulate. We are also interested in applying QLB in other tasks such as electronic voting, online auction and multiparty lotteries.

\newpage




\newpage

\bibliographystyle{spmpsci}
\bibliography{QuantumBlockchain}

\end{document}